\documentclass{ws-ijmpe}

\begin{document}

\markboth{Durier \& de Freitas Pacheco}{Durier \& de Freitas Pacheco}

\catchline{}{}{}{}{}

\title{{\bf THE EVOLUTION OF THE BARYON DISTRIBUTION IN THE UNIVERSE
FROM COSMOLOGICAL SIMULATIONS}}

\author{\footnotesize FABRICE DURIER}
\address{Max Planck Institut for Extraterrestrial Physics, Giessenbachstra$\beta$e~~\\
Garching 85748, Germany,\,
\-email:fdurier@mpe.mpg.de}

\author{\footnotesize J.A. de FREITAS PACHECO}
\address{University of Nice-Sophia Antipolis - Observatoire de la C\^ote d'Azur\\
UMR 6202 - BP 4229, Nice Cedex 4, F-06304, France,\,
\-email:pacheco@oca.eu}

\maketitle

\begin{history}
\received{(received date)}
\revised{(revised date)}
\end{history}

\begin{abstract}

The evolution of the baryon distribution in different phases, derived from cosmological simulations,
are here reported. These computations indicate that presently most
of baryons are in a warm-hot intergalactic (WHIM) medium (about 43\%) while at $z=2.5$ most of baryons constitute the
diffuse medium (about 74\%). Stars and the cold gas in galaxies represent only 14\% of the baryons at $z=0$.
For $z<4$ about a half of the metals are locked into stars while the fraction present in the WHIM and in the
diffuse medium increases with a decreasing redshift. In the redshift range $0\leq z \leq 2.5$, the amount of metals 
in the WHIM increases from 4\% to 22\% while in the diffuse medium it increases from 0.6\% to 4\%. This enrichment 
process is due essentially to a turbulent diffusion mechanism associated to mass motions driven by supernova explosions.
At $z=0$, simulated blue (late type) galaxies show a correlation of the oxygen abundance present in the cold
gas with the luminosity of the considered galaxy that agrees quite well with data derived from HII regions. 

\end{abstract}

\section{Introduction}

Presently, observations suggest that the largest contribution to the energy budget of the universe comes from
unknown forms of matter-energy. ``Standard" matter or baryons represent only a small fraction
of the energy budget. The abundances of elements like $^4He$, $^2D$ and $^7Li$, issued from the primordial
nucleosynthesis, give a direct information about the  baryon-to-photon ratio, a quantity supposed to remain
constant during the (adiabatic) evolution of the universe. Recent analyses of HII regions 
in low-metallicity galaxies$^1$ indicate
a primordial $^4He$ abundance by mass of $Y(He)=0.2565\pm 0.0010$ which, in terms of the critical density, represents
an amount of baryons of $\Omega_b = 0.0492$ for a Hubble parameter equal to $H_0 = 71~kms^{-1}Mpc^{-1}$. Observations
of absorption systems in the spectra of quasars$^2$ point to a deuterium abundance (by number) of
$^2D/H = 2.8\times 10^{-5}$, which can be translated into a density parameter $\Omega_b = 0.0422$. If 
the baryon content resulting
from helium and deuterium data are in quite good agreement, this is not the case for $^7Li$. Lithium abundances
give systematically lower baryon-to-photon ratios, although a recent re-analysis of existing data$^3$ indicates
values close to those already mentioned. Another source of information comes from 
multipole analyses of the cosmic microwave background (CMB) and the fit of the acoustic peaks. Seven years
data$^4$ from WMAP give $\Omega_b = 0.0448$, consistent with values derived from primordial nucleosynthesis.
Finally, the amount of baryons in the universe can also be obtained from the
galaxy-galaxy correlation function derived from different surveys, via the transfer function$^5$. From this method,
the resulting amount of baryons is $\Omega_b = 0.0482$. All these data together 
indicate consistently that
the amount of baryons in the universe is $\Omega_b = 0.046\pm 0.004$.

Once the total amount of baryons is known, a natural question appears: how these baryons are distributed in the universe? 
An early investigation by Fujita, Hogan \& Peebles$^6$ led to the conclusion that at $z\sim 0$ stars 
represent at maximum 12\% of the
total amount of baryons, with the remaining fraction being under the form of cold, warm or hot gas. The cold ($T \leq 10^4$~K) 
and dense gas is present in galaxies, constitutes the reservoir from which new stars are formed, while the cold 
($T \leq 2\times 10^4$ é K), rarefied and photoionized gas forms the diffuse phase pervading the intergalactic medium.
The warm-hot intergalactic medium (WHIM) can be found in different environments: the warm ($T \sim 10^5-10^6$~K) gas
is found in filaments constituting the cosmic web and in halos while the hot gas ($T \sim 10^7-10^8$~K) is found
mainly in the intracluster medium. Damped Lyman absorption features (DLAs) observed in spectra of quasars$^7$
are probably associated to the cold and dense gas. In the one hand, Ly-$\alpha$ absorbers with line-widths 
$b$ less than 40~km$s^{-1}$ (NLAs) observed at low redshifts are thought to be associated with clouds constituted
by a cold photoionized gas present in the IGM and, in the other hand, broad features ($b > 40 kms^{-1}$) or BLAs
are thought to be formed in a warmer medium with temperatures around $T\sim 10^5$~K.  
This picture is supported by high-resolution UV spectra of low redshift quasars obtained with 
the COS spectrograph on board of the Hubble Space Telescope$^9$, which show the presence of O VI absorption features. About 37\%
of the intervening O VI absorbers have velocities centroids coincident with those of the 
HI features. If these absorption lines trace the same gas, then they suggest a temperature $T \leq 10^5\, K$. 
Moreover, 53\% of the O VI features have a complex structure suggesting higher temperatures$^9$. The fraction 
of baryons in these various phases is 
quite uncertain because it depends on some badly known parameters like
the distribution of clouds with the redshift, their physical properties like temperature, density, ionization
degree and metallicity. Different estimates suggest that at low redshifts $30\pm 10$\% of the baryons are in the cold phase 
where DLA features are formed and, about 20\% are in the warm-hot phase traced by the complex and 
broad O VI absoption lines. If the contribution of stars is
taken into account, the amount of "missing" baryons could be of the order of 40\%. Besides baryons mostly in
the form of hydrogen and helium, some heavy baryons or "metals" could also be "missing". This problem, as formulated
by Pettini$^{10}$ many years ago, can be stated as follows. Using the cosmic star formation 
rate (CSFR) derived from UV, optical or radio data and assuming an initial mass function (IMF), the fraction of 
metals at a given redshift can be computed. However, observations indicate that less than 20\% of the expected amount 
of metals is stored in DLAs. Ferrara, Scannapieco and Bergeron$^{11}$ attempted to quantify the extent of the deficit,
suggesting also where the ``missing" metals might be found, i.e., in the hot gas filling galactic halos.

Cosmological simulations are certainly an adequate tool to investigate the evolution of the distribution of baryons (including
metals) in the universe. Using such an approach, it is possible
to probe a significant volume of the universe, to follow consistently the evolution of dark matter and
gas, to test different star formation conditions and to include feedback effects due to massive young
stars (stellar winds and UV ionizing radiation), supernovas (SNs) and AGNs. Past investigations$^{12,13}$
based on such an approach suggested that most of the supposed ``missing" baryons might be in a gaseous phase 
with temperatures in the range $10^5-10^7~K$ (WHIM) and with densities on the average above the background. 
According to these studies, the WHIM is primarily heated by shocks produced during the formation of structures like 
filaments by gravitational instability, with the feedback due to SNs playing a secondary role$^{13}$. 
These simulations$^{12}$ have also shown that there is a steady transfer of gas from the diffuse IGM to the WHIM 
mainly for $z \leq 3$. According to the simulations by Oppenheimer \& Dav\'e$^{14}$, the fraction of metals
present in the WHIM phase is practically constant in the context of their prescription for galactic winds
($\sigma$-wind model), whereas simulations based on a different prescription$^{15}$ ($\phi$-wind model) indicates
an increasing fraction from $z \leq 4$, with about 25\% of metals being in this phase at the present time.

In this work we report results from cosmological simulations performed
with the Nice cosmological code, aiming to follow the evolution of the 
baryon distribution in different gaseous phases and stars 
as well as their chemical enrichment. Our simulations indicate that presently about 14\% of the baryons are locked into
stars, 7\% are in the form of a cold and dense gas in galaxies, 36\% constitute the diffuse medium and the
remaining 43\% are in the WHIM. For $z < 4$ about half of metals are locked into stars. 
The paper is organized as follows: in Section 2 the simulations are briefly described, in Section 3
the main results are reported and finally, in Section 4 the main conclusions are given.

\section{The simulations}

Simulations were performed by using the parallel TreePM-SPM code GADGET-2 in a formulation that conserves energy
and entropy$^{16}$. This code is fully adaptive in space and time, permitting simulations with an adequate
dynamic range required to study both high (galaxies and halos) and low (IGM) density regions.
Different physical mechanisms affecting the dynamics and the thermodynamics of the gas were
included, such as cooling (collisional excitation of HI, HeI and HeII levels, free-free transitions, radiative 
recombinations, $H_2$ transitions, atomic fine-structure level exitations of trace elements and 
Compton interactions with CMB photons), local heating and ionization
by the UV-radiation of newly formed stars, mechanical energy injected either by type II and type Ia SNs or
AGNs. 

Some particular aspects of the Nice code should be emphasized. The first is that the return of mass to the interstellar
medium, consequence of the stellar evolution, is included and, consequently, the mass of a gas-particle
varies in time. The second point is that the delay between the appearance of type II and type Ia SNs is
taken into account. The onset of type Ia SNe occurs about 800 Myr after the beginning of a star 
formation episode, i.e., after the conversion of a gas-particle into a stellar-particle. Then, the
probability to occur a type Ia event is calculated according to the prescription by Idiart et al.$^{17}$.
The third aspect to be strengthened concerns the chemical enrichment
and the diffusion of metals. Past studies$^{18}$ assumed that metals ejected by SNs are redistributed
according to the usual SPH kernel interpolation technique. An algorithm adequate for SPH simulations and able to
describe a turbulent diffusion process of metals was introduced by Greif et al.$^{19}$. A similar approach
was adopted in our simulations. Each stellar-particle is considered as a ``diffusive" center and, in this case,
for a constant diffusion coefficient $D$, the concentration $Z$ of a given element varies as
\begin{equation}
\frac{\partial Z}{\partial t}=D\nabla^2Z
\end{equation}
The solution of this equation was simulated by considering that the concentration of metals in a given gas-particle
depends on the concentration difference with respect to neighbors. Since the diffusive flux is
zero when no concentration gradients exist, the mean concentration ${\bar Z}$ of all particles inside
the cell is calculated as well as the quantity
\begin{equation}
Z_N(i)=\sum_{j}\frac{(Z_i-Z_j)}{r^2_{ij}}
\end{equation}
where $r_{ij}$ is the distance between the gas-particles $i$ and $j$.
Then, each gas-particle in the SPH cell will receive an amount of heavy elements given by
\begin{equation}
\Delta Z_j=\eta (Z_i-{\bar Z})\frac{(Z_i-Z_j)}{Z_n(i)r^2_{ij}}
\end{equation}
The parameter $\eta\sim (0.01-0.02)$ is essentially the ratio between the squares of the diffusion length and the
dimension of the cell, i.e., $\eta\sim D\Delta t/L^2$. 
Concerning the injection of mechanical energy by AGNs in the surrounding medium, two injection-modes were
considered$^{20}$. In the first (``disk mode"), when the supermassive black hole (SMBH) is in a state of accretion, it is
supposed that 10\% of the disk bolometric luminosity is injected in the medium through two opposite jets
having an aperture angle of $20^o$ and a length of about 300~kpc. In the second (``Kerr mode") the energy rate
injected in the medium by the jets is given by the relation$^{21}$
\begin{equation}
L_{jet}=4\times 10^{28}\left(\frac{H}{10^4~G}\right)^2M_{bh}^2 ~~erg/s
\end{equation}
where $M_{bh}$ in solar units is the mass of the SMBH. The numerical constant was obtained by assuming
that the average spin of SMBHs in geometric units is $S=0.46$.

All simulations were performed in a cube with a size of $50h^{-1}$ Mpc and a flat $\Lambda$CDM cosmology was
adopted, characterized by a Hubble parameter $H_0=70~kms^{-1}Mpc^{-1}$, a matter density parameter 
$\Omega_m=0.3$ and ``vacuum" density parameter $\Omega_{\Lambda}=0.7$. The baryonic density parameter
was taken to be $\Omega_b=0.04$ and the normalization of the matter density fluctuation spectrum was taken to
be $\sigma_8=0.9$. Simulations were performed with two different mass resolutions for gas-particles: 
$4.84\times 10^8~M_{\odot}$ (four runs) and $1.18\times 10^8~M_{\odot}$ (two runs). 
Effects of the feedback by SNs were investigated in runs with lower mass resolution
while the two energy injection modes by AGNs where studied in the two runs having a better mass resolution. 
Computations were performed at the {\it Center of the Numerical Computation of 
the C\^ote d'Azur Observatory} (SIGAMM).

\section{Results}

Low resolution runs were performed to explore feedback effects due to supernovas. No dramatic differences in the
baryon fraction present in different gaseous phases were noticed for the explored range of injected energies,
except for one extreme case. In fig.~1 is shown the baryon fraction in the cold phase for two runs
in which the injected energy by type II SN in the ISM varied from $\alpha_{II}$=0.03 to $\alpha_{II}$=0.07 
and the injected energy by type Ia SN was kept constant and equal to $\alpha_{Ia}$=0.07 (all these values are
in units of $10^{51}~erg$). As it can be seen, increasing the injected energy by a factor 2.5, the fraction of cold gas decreases 
on the average by 17\%. This transfer effect (cold to WHIM phase) increases with increasing redshift, corresponding  
to a reduction in the cold gas 
fraction of about 4.0\% at $z=0$ and of about 43.0\% between $z\sim$ 2.5-3.0. We have also performed a run with
$\alpha_{II}$=0.20, a highly unrealistic value. In this case, the rapid appearance of type II events and the
large amount of injected energy in the ISM affect the local formation of new stars and their chemical enrichment.
New stellar generations are no more formed in this newly metal enriched environment but in other places in which
more favourable conditions are present. These regions have an adequate gas temperature and density but are
in general less metal rich. Consequently, the evolution of the mean stellar metallicity resulting from this run displays 
values at a given redshift lower than the precedent experiments. 

\begin{figure}[th]
\centerline{\psfig{file=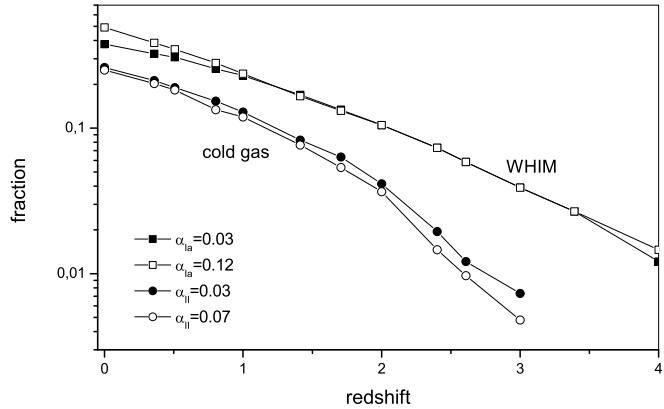,width=10cm}}
\vspace*{8pt}
\caption{Evolution of the cold gas fraction for two different values of the injected energy in ISM by type II
supernovas as well as the evolution of the gas fraction in the WHIM phase when the injected energy by type Ia
supernovas is varied. Labels give the feedback energy in units of $10^{51}$~erg.}
\end{figure}

This is not the case when the injected energy by type Ia SN is varied. Runs with the feedback energy by type Ia SN 
being respectively equal to $\alpha_{Ia}=$0.03, 0.07 and 0.12 do not indicate any noticeable effect in the
evolution of the mean stellar metallicity. This can be easily understood since type Ia events occur only 800 Myr after
the onset of the star formation in a given place and, in the meantime, the next generation of stars can be formed from
a cold gas enriched in heavy elements and only slightly affected by the feedback of type II SN.
If no significant effects were seen in the evolution of the mean stellar metallicity, the fraction of baryons in the
WHIM phase increases as the injected energy by type Ia SN increases. Figure 1 shows the evolution of baryons in the WHIM phase
for two values of the injected energy in the ISM by type Ia events ($\alpha_{Ia}$=0.03 and $\alpha_{Ia}$=0.12). Differences
appear around $z\sim$1.5, when the maximum of the frequency of type Ia SN occurs. At z=0, the fraction of baryons
in the WHIM phase increases by about 30\% when the feedback energy passes from $\alpha_{Ia}$=0.03 to $\alpha_{Ia}$=0.12.
More details concerning specifically the injection of thermal and mechanical energy by supernovas 
can be found elsewhere$^{22}$. 

\begin{figure}[th]
\centerline{\psfig{file=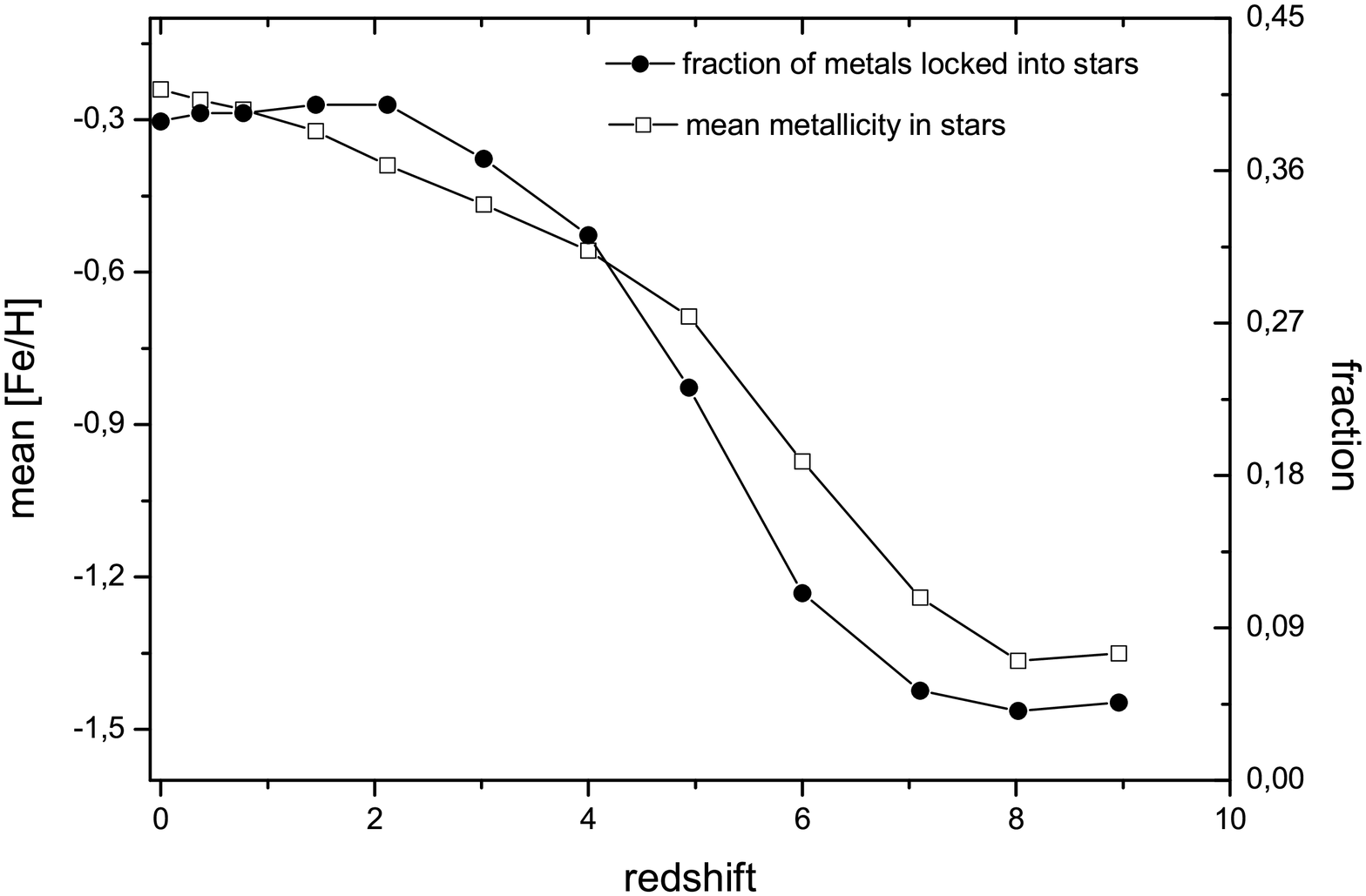,width=10cm}}
\vspace*{8pt}
\caption{Evolution of the mean stellar iron abundance and of the fraction of metals locked into stars
derived from a run using the AGN feedback in the ``Kerr mode". The feedback by SNs correspond
respectively to injected energies $\alpha_{II}=0.05$ and $\alpha_{Ia}=0.01$.}
\end{figure}

From the analysis of the two runs with better mass resolution, we concluded that there is no significant differences 
either in the baryon distribution or in the metal enrichment history, when the adopted feedback of AGNs is 
the ``disk mode" or the ``Kerr mode". Thus,
the results reported here refer basically to the ``Kerr mode". In fig.~2 are shown both the evolution of the mean
stellar iron abundance and the fraction of all produced metals locked into stars at a given redshift. Notice that
in the redshift range $8 > z > 2$ the mean stellar metallicity increases by one order of magnitude, being presently
about 57\% of the solar value. On the other hand, for $z < 3$, the fraction of all produced metals locked into stars varies
very little, remaining around 40\%. These results agree quite well with those of Dav\'e \& Oppenheimer$^{15}$, whose
simulations were performed with the same number of particles and similar mass resolution but disagree with
the simulations by Shen et al.$^{23}$, who concluded that presently about 82\% of metals are locked into stars.

Fig.~3 shows the evolution of baryons in different phases. At redshifts $z > 4$ the gas is mostly in the
diffuse phase (more than 90\%). As structures are formed by gravitational instability, the amount of cold gas 
present in assembling galaxies increases, reaching a maximum ($\sim$~18\%) around $z\sim$ 1.5
and then decreases due essentially to the star formation process and also from
the passage to the warm-hot phase due to feedback processes.
Notice that the fraction of shocked gas constituting the WHIM increases continuously and 
this phase stores presently most of baryons ($\sim$ 43\%). The fraction of baryons in stars
is also an increasing function, representing today about 14\% of all baryons in the universe. This is
comparable with the simulations by Rasera \& Teyssier$^{24}$, who found that the baryon fraction under
the form of stars at $z=0$ is $\sim$~12\%. However, these authors concluded that presently only 29\% of
the baryons constitute the WHIM and most of them are still in the diffuse medium (58\%). 

\begin{figure}[th]
\centerline{\psfig{file=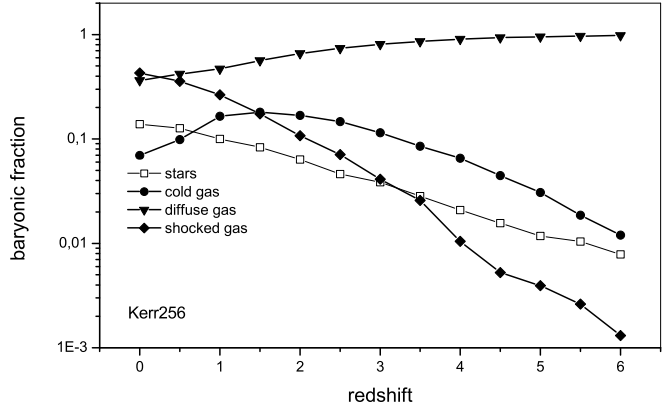,width=10cm}}
\vspace*{8pt}
\caption{Evolution of the fraction of baryons in different phases derived from
a run using the AGN feedback in the ``Kerr mode". The feedback by SNs correspond
respectively to injected energies $\alpha_{II}=0.05$ and $\alpha_{Ia}=0.01$}
\end{figure}

The evolution of the fraction of metals present in the different gaseous phases and stars is shown in fig.~4. 
Since stars are formed from the cold gas and inject their nucleosynthesis products primarily in this phase , the
amount of metals in stars and in the cold gas are always comparable. In the WHIM and in the diffuse gas, the amount
of metals increases with time but is considerably less than the fraction present in the cold dense medium and stars.
This behavior can be understood by the superposition of two effects: the (turbulent) diffusion mechanism and
the mass-flows driven by the mechanical energy injected by SNs, which spread out the metals present 
in the cold gas, where they were initially injected, to the WHIM and from there to the diffuse medium.

\begin{figure}[th]
\centerline{\psfig{file=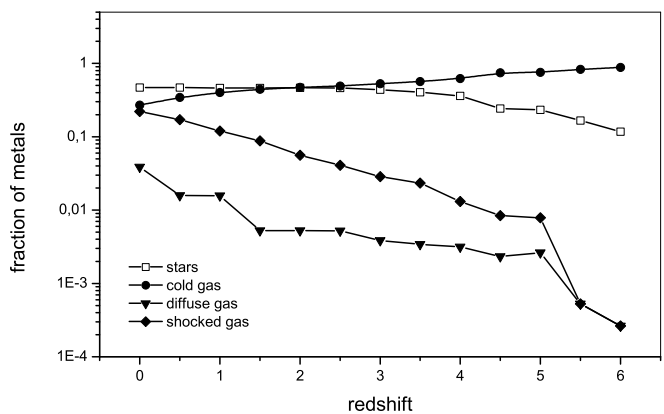,width=10cm}}
\vspace*{8pt}
\caption{Evolution of the fraction of metals present in different phases
derived from a run using the AGN feedback in the ``Kerr mode". The injected
energy by SNs are respectively $\alpha_{II}=0.05$ and $\alpha_{Ia}=0.01$.}
\end{figure}

The cold dense gas is probably related with the medium where DLA features are formed. In fig.~5 the evolution
of the mean metallicity in this phase derived from simulation using the ``Kerr-mode" for AGNs
is shown. For comparison, data from Kulkarni et al.$^{25}$ and Prochaska et al.$^{26}$ are also shown. Notice 
that simulated ``metallicities" refer to iron abundances while data on DLAs refer to zinc and despite the 
fact that both elements are referred to solar values, the abundance pattern at $z > 1$ is certainly not solar and 
some small differences are expected in the behavior of both elements. Even so, the agreement between  simulations
and data is quite satisfactory. 

\begin{figure}[th]
\centerline{\psfig{file=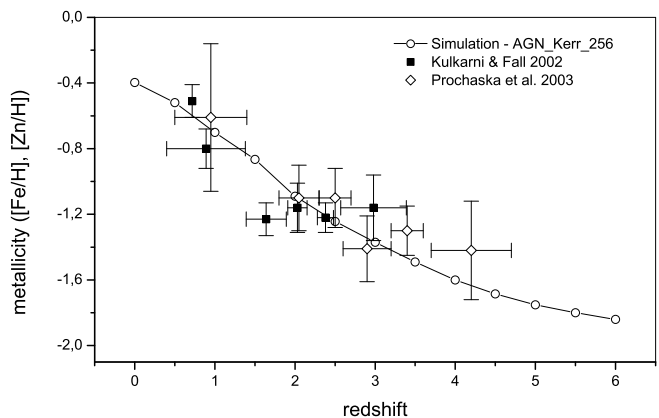,width=10cm}}
\vspace*{8pt}
\caption{Evolution of the metallicity in the cold dense gas phase derived from the run using the ``Kerr mode" for 
the AGN feedback. Data points on DLAs refer to metallicities derived from Zn data.}
\end{figure}

In the catalog of our simulated galaxies, two main types are identified: the ``red" galaxies, objects having colors redder 
than $(U-V)>1.1$ and $(B-V)>0.8$, and the others, classified as ``blue" galaxies. The former have, on the average,
higher metallicities and a mean stellar population age of about 8.5~Gyr while the latter have a younger stellar
population (mean age of $\sim$ 5.5~Gyr), reflecting the occurrence of more recent star formation episodes.
Blue galaxies still retain presently a non-negligible amount of cold gas, rich in metals. Young stars ionize
such a gas and abundances can be derived from emission lines formed in the corresponding HII regions. In fig.~6
the oxygen abundances of such a cold gas, gravitationally bound to ``blue" galaxies, derived from the
simulation AGN-Kerr is plotted against the blue luminosity of corresponding galaxies. There is clearly
a luminosity-metallicity (or mass-metallicity) relation, with brighter galaxies being more metal rich. Notice
that for resolution reasons, no objects fainter than $M_B \sim$ -18.5 are present in our plot. Mean oxygen
abundances derived from HII regions belonging to galaxies of the local universe are also shown for comparison.

\begin{figure}[th]
\centerline{\psfig{file=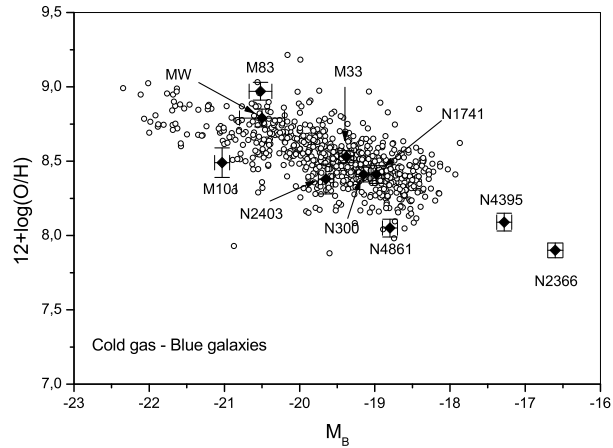,width=10cm}}
\vspace*{8pt}
\caption{Oxygen abundances for simulated ``blue galaxies" (open dots) as a function of the absolute
B-magnitude. Data points correspond to mean oxygen abundance values derived from HII regions in the
corresponding galaxy. Error bars indicate dispersion around mean values.}
\end{figure}

The baryon and the metal budget derived from our simulation AGN-Kerr at two reference redshifts is summarized in
table~1. Almost of 70\% of baryons are presently in warm-hot intergalactic medium and in the cold
phase, while the remaining fraction is shared between stars and the diffuse medium. The situation is
rather different at $z = 2.5$ where baryons are essentialy in the cold and diffuse phases with only
a small fraction being under the form of stars and in the WHIM. Concerning the metals produced by 
nucleosynthesis in stellar interiors, they can be found predominantly in stars and in the cold phase.
The fraction present in stars does not vary significantly in the redshift range $0-4$. On the contrary,
the fraction of metals in the cold phase decreases steadly due to diffusion, increasing the contamination
of the WHIM and the diffuse medium.

\begin{table}
\tbl{Baryon and metallicity budget at two reference redshifts derived from the simulation AGN-Kerr}
{\begin{tabular}{cccccc}
\toprule
 & Baryon&Budget& &Metal&Budget \\
\colrule
phase&z = 0&z = 2.5& &z = 0& z = 2.5\\
\colrule
stars & 13.8\% & 4.6\%& & 47.0\%&46.1\%\\
cold gas & 7.0\% & 14.7\%& &27.0\%&49.2\%\\
WHIM & 42.8\% & 7.0\%& & 22.1\%&4.1\%\\
diffuse medium &36.4\% & 73.7\%& & 3.9\%&0.6\%\\
\botrule
\end{tabular}
\label{tb1}}
\end{table}

\section{Conclusions}

Cosmological simulations are certainly one of the best tools to investigate the
evolution of the different phases (gaseous and stellar) in which baryons are distributed
in the universe. Such a numerical approach is useful to study the chemical
enrichment not only of stars, but also of the nearby gas where metals are injected
by SN explosions and that of the intergalactic medium by diffusion processes.

Despite the necessity of improving the treatment of the feedback processes, the present
investigation indicates that type Ia SNs, depending on the adopted value for the injected
mechanical energy, contribute to increase the fraction of baryons on the WHIM for $z < 1.5$
at the expense of the cold dense gas.

The present fraction of baryons in the form of stars, cold gas, warm-hot intergalactic medium
and diffuse medium resulting from the present simulations is consistent with estimates
derived from the present data. In particular, the evolution of the mean metallicity of the 
cold gas is consistent with abundances of DLA features. The present oxygen abundances of the
cold gas gravitationally bound to ``blue" (late-type) galaxies, derived from our simulations,
indicate the existence of a luminosity-metallicity (mass-metallicity) relation and
are in good agreement with oxygen abundance data of HII regions in galaxies of the local universe.

\end{document}